# Determining Factors for the Accuracy of DMRG in Chemistry

Sebastian F. Keller and Markus Reiher*

*Correspondence*: Prof. Dr. Markus Reiher

ETH Zürich

Laboratorium für Physikalische Chemie

Vladimir-Prelog-Weg 2

CH-8093 Zürich

Tel.: +41 44 633 4308

Fax: + 41 44 633 1594

E-mail: markus.reiher@phys.chem.ethz.ch

*Abstract:* The Density Matrix Renormalization Group (DMRG) algorithm has been a rising star for the accurate ab initio exploration of Born-Oppenheimer potential energy surfaces in theoretical chemistry. However, owing to its iterative numerical nature, pit falls, that can affect the accuracy of DMRG energies, need to be circumvented. Here, after a brief introduction into this quantum chemical method, we discuss criteria that determine the accuracy of DMRG calculations

## 1. Introduction

The Born-Oppenheimer approximation provides two central ingredients for chemistry: one is the concept of molecular structure [1] and the other one is the electronic energy, which acts as a potential energy that determines the motion of the atomic nuclei (in this approximation the 'potential in which the nuclei move') and from which thermodynamic as well as kinetic insights can be extracted. The electronic energy is thus central for theoretical chemistry and calculated as the eigenvalue of the electronic Schrödinger equation. However, the accurate solution of this equation is a delicate problem. Two major directions emerged, namely wave-function theory (WFT) and density-functional theory (DFT). While DFT is doubtlessly the most frequently applied approach in quantum chemistry, it lacks the option of a systematic improvement on results obtained with some setting (i.e., with some choice for an approximate functional). For this reason, wave-function methods are under continuous development, although they are usually prohibitively expensive in terms of computer time for large molecules of, say, 100 or more atoms.

The standard ansatz in these latter methods is the pre-definition of a many-electron basis set. From a chemist's perspective, this may be viewed as a generalization of the textbook LCAO concept for orbitals to the total electronic wave function, which is the eigenfunction in the electronic Schrödinger equation. The standard many-electron basis set comprises (linear combinations of) Slater determinants containing the molecular orbitals. The linear expansion parameters in front of the determinants can be determined either variationally (e.g., in configuration interaction (CI) methods) or by projection (as performed in standard coupled-cluster (CC) approaches). Although the expansion in terms of Slater determinants assumes an independent-particle picture (as it rests on the orbital approximation), it can be made exact, if the determinant basis is complete. Then, any electronic state of a molecule can be expanded exactly in such a complete many-electron basis set, which is called 'Full-CI' in chemistry and 'exact diagonalization' in physics. Unfortunately, the albeit simple construction of this complete basis set comes with the flaw that the basis-set size grows factorially, which makes its construction by a computer program unfeasible but for the smallest molecules.

Naturally, approximations have been devised which contributed popular and accurate methods like multi-reference CI or singles and doubles plus perturbatively treated triples coupled-cluster, CCSD(T), to the tools of trade of computational chemistry. Although the success of CC models is remarkable and although highly efficient implementations have been devised [2], their extension to the general multi-configurational case turned out to be cumbersome so that no clear-cut, efficient solution of this problem appears to be in sight. However, from a different field, namely the physics of spin chains, a totally new ansatz emerged: the Density Matrix Renormalization Group (DMRG). The DMRG algorithm was designed for correlated quantum problems in Condensed Matter Physics [3-5], which usually assume a local, nearest-neighbors-only interaction operator. Despite the fact that the full Coulomb interaction of the electrons in a molecule does, in general, not sustain any of the locality assumptions made for the development of DMRG, it was shown that DMRG can be applied to challenging quantum-chemical problems [6-9]. A formal analysis of the electronic state optimized by DMRG is a so-called 'matrix product state' (MPS) [10]. The MPS concept can be generalized to a more general framework, which has been called 'tensor network state'

[11] and has also found application for the full quantum chemical interaction operator [12-14].

For a given (finite) set of molecular orbitals, DMRG can systematically approach the finite-basis Full-CI result in this active orbital space. DMRG is variational and also capable of describing multi-reference states occurring in complicated electronic situation as found, for instance, in transition metal complex and cluster chemistry [15]. However, there is a catch because the accuracy of a DMRG calculation depends on a set of determining factors. While the role of these factors has been fully appreciated by the experts in the field, here we shall provide an overview of them that may help establish DMRG as a standard technique of the toolbox of the computational chemist.

## 2. Theoretical background

Similar to a CI wavefunction ansatz, the electronic DMRG wave function can be respresented as a linear combination of Slater determinants constructed from $L$ spatial orbitals

$$|\psi\rangle = \sum_{\boldsymbol{\sigma}} M^{\sigma_1} M^{\sigma_2} \cdots M^{\sigma_L} |\boldsymbol{\sigma}\rangle,$$

*Equation 1: In Matrix Product States, expansion coefficients of Slater determinants are encoded as products of matrices.*

in which the coefficients are encoded as a product of matrices $M^{\sigma_i}$. Therefore, states in this particular format are called Matrix Product States (MPS). The bold index $\boldsymbol{\sigma}$ is an abbreviation for the the occupation number vector ($\sigma_1, \ldots, \sigma_L$), which runs over all possible orbital occupations and thus labels the orthonormal basis states of the $L$-orbital system that are the Slater determinants. For each orbital, there is a set of four matrices corresponding to the four possible orbital occupations, labeled by the upper index $\sigma_i$. Choosing an occupation for each orbital contributes to a Slater determinant and determines one matrix per orbital, in DMRG jargon called *site*. The contraction of the selected matrices via ordinary matrix-matrix multiplication yields the CI coefficient of the corresponding determinant [16,17]. Since these coefficients are scalar, the matrices for the first and last orbitals are required to be row and column vectors respectively.

The DMRG algorithm then consists of an iterative protocol, in which the site matrices are variationally optimized with respect to the energy in sequential order. The basic ingredient of these local optimization problems is the diagonalization of the matching local part of the electronic Hamiltonian operator, which can be achieved by employing a sparse diagonalizer such as the Jacobi-Davidson algorithm. The sites that are undergoing optimization constitute the active subsystem, as shown in Figure 1. As a result of the optimization, the entries of the site matrices are replaced by a new, optimized set of entries which correspond to the eigenvector of the local Hamiltonian operator with the lowest eigenvalue. Combining the local optimizations, the electronic ground-state energy is calculated iteratively by passing through all sites from left to right and vice versa, referred to as a *sweeping*, until the energy is converged. The rate at which this happens strongly depends on whether a single site is optimized at a time (single-site DMRG) or two sites are simultaneously optimized (two-site DMRG). While the former variant performs well for local interaction operators, such as different versions of the Hubbard model in Solid State Physics, in Quantum Chemistry this single-site DMRG gets trapped in local energy minima for even the simplest systems. Two-site DMRG turns out to have much more robust convergence properties, albeit at a higher numerical cost.

A central feature of the ansatz in Eq. (1) is that its precision can be controlled by adjusting the maximum dimension $m$ that each matrix is allowed to assume. If $m$ is allowed to grow exponentially according to the Hilbert space dimension of a system, DMRG becomes essentially the Full-CI method. What allows DMRG in many cases to perform much better in terms of the scaling behavior than standard Full-CI, however, is that the maximum matrix dimension $m$ needs only account for a tiny fraction of the exponentially large Hilbert space to achieve the same result as Full-CI within numerical precision. In one dimensional systems with finite-range interactions, $m$ can even be held constant with increasing system size and without loss of accuracy. The physical basis for this phenomenon, the so-called area laws of entanglement, has been rigorously studied by the Condensed Matter Physics and Quantum Information Theory community [18, 19]. The amount of entanglement between the two parts of any system bipartion, measured by the von Neumann entropy, is either constant or a logarithmic function of the system size [20]. Because the maximum amount of entanglement that a state in the form of Eq. (1) can encode is determined by $m$, this parameter is a central quantity of the DMRG algorithm. It is called the number of 'renormalized basis states' or the number of 'block states' as it is also equal to the number of left and to the number of right subsystem basis states, respectively, if one divides the total system into two parts, see Figure 1. It is important to note that the number of many-electron DMRG basis states, each of which can be understood as complicated linear combinations of Slater determinants, is actually $m \times 4 \times 4 \times m = 16m^2$ for the two-site DMRG algorithm. Hence, the number of variational parameters, which are CI-type expansion coefficients in front of the DMRG basis states, is also $16m^2$.

Another appealing feature of DMRG is that, at least in the two-site variant of the algorithm, the error introduced by limiting $m$ can be tracked in a systematic way which will become clear in a moment.

Extending the left and the right subsystem of orbitals by the adjacent active site on its right and on its left, respectively, yields the total system now bipartitioned into two subsystems represented by $4m$ many-electron basis states. The eigenvalue problem for the combination of both enlarged subsystems is thus of dimension $16m^2$. After diagonalizing the Hamiltonian of this total system – sometimes called the superblock Hamiltonian – a reduced density matrix can be constructed from the Hamiltonian eigenvector by tracing out all states on the complementary subsystem. Diagonalizing this reduced density matrix for the active subsystem yields *m* eigenvectors with highest eigenvalue that form a rectangular renormalization matrix needed for the dimension reduction of all creation and annihilation operators in the Hamiltonian from *4m* back to *m*. The truncated weight of the *m* states, defined as the sum of their eigenvalues, is a useful measure for the accuracy of the approximate wave function as it tends to zero if *m* is increased towards the dimension of the complete Hilbert space. The selection of *m* highest-eigenvalue eigenvectors for the dimension reduction of all operators (called 'decimation') can be understood as a least-squares fit to reduced-dimensional many-electron basis sets defined for the two subsystems.

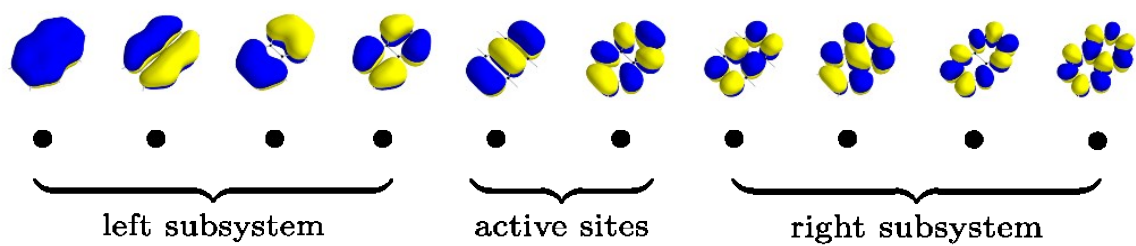

*Figure 1: Partition of a chosen active molecular orbital space into left, active and right subsystems on which DMRG many-electron basis states are constructed. If the sweep is processed from left to right then the left subsystem is the active subsystem, on which many-electron states are systematically constructed, while the right (complementary) subsystem carries the many-particle states optimized in the previous sweep processed in opposite direction, i.e. from right to left. In a DMRG iteration step, the left subsystem is now enlarged to incorporate the left active site, while the right active site is absorbed into the right subsystem.*

What makes DMRG a successful method in Quantum Chemistry is that between $10^3$ and $10^4$ subsystem states *m* are usually sufficient to reduce the truncated weight enough to calculate ground state energies with sub-mHartree precision.

### 3. Performance of the method

As with other active-space methods, the decisive parameter determining the cost of a calculation is the number of correlated electrons in the number of active orbitals. While traditional methods like the complete-active-space self-consistent-field (CASSCF) approach are exponentially expensive with respect to the size of the active space, DMRG formally exhibits a scaling of $O(m^2L^4) + O(m^3L^3)$. One must keep in mind, however, that this behavior breaks down if the amount of entanglement measured by the von Neumann entropy between the two parts of any bipartition of the system increases with the system size. If this is the case, the number of renormalized basis states $m$ needed to attain a certain accuracy may increase dramatically. In the worst case, exponential scaling is recovered. Fortunately, many molecules with a complicated electronic structure still exhibit some degree of locality such that the amount of entanglement does not increase linearly with system size.

Besides the fundamental limitations of active space size and entanglement, other, more technical aspects also have a strong influence on DMRG convergence and thus on efficiency and possibly also on accuracy. These issues shall be discussed in some depth in the next section.

**4. Determining factors of DMRG convergence and accurracy**

In this section, we provide an overview of determining factors that affect DMRG accuracy and are thus essential factors to report if results from DMRG calculations are exploited in chemical research.

The example data provided in this section were computed with our new, massively parallel MPS-based quantum-chemical DMRG program QC-MAQUIS [21] and with the Budapest quantum-chemical DMRG program developed by Ö. Legeza since 2000 [22].

**4.1 Size of the active space**

Despite the substantially higher computational efficiency that DMRG achieves compared to Full-CI, the exact treatment of the electron correlation problem in the full orbital basis remains out of reach on today's computers. For this reason, the concept of a complete active space, which emerged during the developement of traditional electron correlation methods, is important for DMRG calculations as well. This concept selects a limited part of the full orbital basis, in which the electronic correlation is treated exactly and which is chosen such that it presumably contains the relevant contributions of the differential electron correlation decisive for chemical processes. Physically, this subspace can be selected in a meaningful way, because the strongest correlation effects occur within the orbitals that lie, in the language of physics, close to the Fermi level (in chemical terms, these are the frontier orbitals). The rest of the full orbital basis below and above the active space are referred to as 'core' and 'virtual' orbitals, and they are assumed to be fully occupied and empty, respectively. The number of determinants involving virtual orbitals neglected from the active space may contribute significantly to the total electron-correlation energy. This type of correlation is referred to as dynamic correlation. The advantage of such a subdivision of the total orbital basis is that dynamic correlation effects can be accounted for by perturbation theory or other methods [23, 24] Often, the choice of a subdivision is made based on empirical criteria. As a tool to aid choosing an appropriate active space and assessing its reliability, we have developed a set of criteria based on entaglement measures [25,26]

**4.2 Choice of the molecular orbitals**

The introduction of an active space entails that only a part of the orbital basis is contained within the model. These molecular orbitals are therefore well suited for a calculation if they allow one to express a large share of the total static correlation within an active space of a computationally feasible size. Hence, the electronic energies depend on the orbital choice. If an active space smaller than the one intended to be used for the DMRG calculation can be meaningfully selected in a preceding CASSCF calculation, the resulting preoptimized CASSCF natural molecular orbitals will be better suited for the subsequent DMRG treatment than plain Hartree-Fock molecular orbitals.

Employing localized molecular orbitals can have an effect on the performance of DMRG [27]. If long-ranged 'interactions' among localized orbitals become weak enough, it might be possible to eliminate some of the terms in the electronic Hamiltonian by a screening of the two electron integrals without affecting accuracy. If localized molecular orbitals in addition help to reduce the amount of entanglement in the system, a smaller number of renormalized basis states will be required to attain a certain accuracy, such that the performance is further enhanced.

The orbital-dependence of DMRG energies may be conveniently resolved by combining the DMRG optimization with an orbital relaxation protocol [28, 29] as implemented in standard methods like CASSCF. Then, an optimum orbital set for some chosen active orbital space may be found that minimized the DMRG electronic energy.

### 4.3 Environment-state guess in the first sweep

DMRG is an iterative method. The speed at which a converged solution is obtained strongly depends on the initial guess, i.e. the initial content of the $M$ matrices in Eq. (1). In practice this guess requires the explicit construction of many-electron states in the complementary subsystem in the first (warm-up) sweep [27,30,31]. The simplest option is to encode the Hartree-Fock determinant and to add white noise to the reduced density matrix in order to avoid losing important basis states in the warm-up sweep [27]. Instead of white noise, one may apply a perturbative correction [32]. However, a more involved alternative, the CI-DEAS protocol [31] may achieve the fastest convergence, i.e. requires the smallest number of sweeps. In the warm-up sweep, the $m$ most important determinants of the complementary subsystem (the environment) are selected such that the entanglement between it and the rest of the system is maximized. If the first sweep starts at the leftmost site, the environment corresponds to the right subsystem in Figure 1 during the first sweep. Table 1 shows how two different initial guesses affect the convergence rate in ground-state calculations of the $F_2$ molecule with 14 electrons in 32 active orbitals and $m=1024$.

| Warm-up guess | Hartree-Fock | CI-DEAS |
|---|---|---|
| Sweep 1 | -198.886421 | -198.962112 |
| Sweep 20 (converged) | -198.970559 | -198.970711 |

Table 1: Total DMRG electronic energy of $F_2$( with a CAS14,32) converged starting from a Hartree-Fock (middle column) and a CI-DEAS initial state (right column) All calculations were carried out with m=1024 and an internuclear distance of 141 pm. Pre-optimized natural orbitals from a CASSCF(14,8) calculation were employed. The orbital orderings for the Hartree-Fock guess was the energetical ordering of the orbitals, while the mutual information had been exploited for the optimized of the ordering for the CI-DEAS guess [33,35,36]. Note that for the moderate choice of m=1024, the converged energies still depend on the initial guess.

### 4.4 Ordering of orbitals

The ordering of the one-dimensional chain of orbitals can also have a strong impact on the convergence characteristics of a DMRG calculation [27,32,34]. Unfortunately, there exists no obvious or simple relation between the ordering, which determines which explicitly treated site is considered next in a sweep, and the convergence properties of the iterations. Moreover, the choices of an optimal ordering and initial guess are intertwined. While a Hartree-Fock initial state performs well in combination with an energetical orbital ordering, the entropy based CI-DEAS initial guess develops its full potential in combination with an ordering that tends to group pairs of orbitals with large mutual interaction close to each other [33, 35,36].

### 4.5 Choice of the number of renormalized subsystem states m

The magnitude of the error introduced in DMRG by limiting the number of renormalized basis states to some fixed number $m$ may limit the accuracy of a converged total electronic energy. However, $m$ is chosen by feasibility constraints as the computational effort depends on $m$. Whether or not a chosen value for $m$ might introduce errors can be assessed by inspecting the eigenvalue spectrum of the reduced density matrix as this determines the optimum number $m$ of eigenvectors to be selected for the decimation step. In practice, calculations with different values for $m$ may be extrapolated [27, 36, 37].

$$\ln\left(\frac{E_{\text{DMRG}} - E_{\text{FCI}}}{E_{\text{FCI}}}\right) = a\ln(\varepsilon) + b$$

*Equation 2: Fit function to extrapolate the truncation error ε towards zero.*

Because the truncation error ε tends to zero if *m* is increased, the precision of the obtained energies can be estimated by extrapolating them towards a truncation error of zero, which corresponds to an unlimited ressource of *m*. A possible fit function, Eq. (2), which works well if *m* is varied over an order of magnitude or more, was given by Legeza et al. [30]. In Figure 2, an extrapolation employing this fit function was performed for different ground state calculations of the $F_2$ molecule. From the figure, the estimated error is less than 10 microHartree.

Besides extrapolation, the energy fluctuations that occur in two-site DMRG when the wave function is almost converged also indicate, whether the number of renormalized basis states was chosen sufficiently large. With increasing *m*, the difference between the minimum and maximum energy obtained within a sweep tends to zero. The reason for these fluctuations is that the two explicitly treated sites in between the two subsystems can combine to produce different amounts of entanglement, depending on their position in the orbital chain. If *m* was not limited, no loss of entanglement would occur and thus no fluctuations either. The error in the energy is therefore at least as large as the fluctuations.

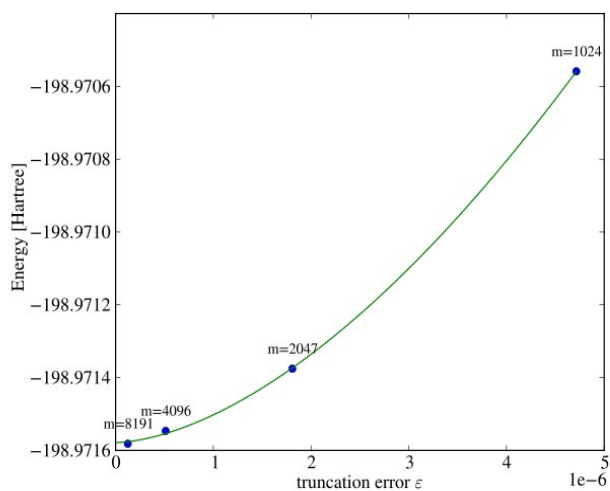

Figure 2: $F_2$ ground-state energies for different m as a function of the truncation error.

## 5. Conclusions

Different parameters of the DMRG algorithm determine the accuracy and performance DMRG calculations. As with other active space methods, the selection of the orbital basis and the active space has a strong effect on the quality of results obtained. However, by contrast to standard approaches like CASSCF, DMRG involves another decisive parameter and that is the number of renormalized states kept on the active subsystem during the iterations. In addition, the factors that control DMRG convergence are the initial guess and the orbital ordering. Finally, the accuracy of results can be verified by performing an extrapolation. All these parameters need to be reported and kept in mind when DMRG results are discussed. Hence, the favorable polynomial scaling of DMRG, which makes it a true competitor to standard methods, affords a set of parameters that need to be well controlled in order to guarantee results exploitable in the study of spectroscopy and chemical reactivity.

**Acknowledgements**

This work was supported by ETH Research Grant ETH-34 12-2.